\documentclass[12pt]{article}
\usepackage{amsmath,amssymb,amsthm}
\usepackage{graphicx,psfrag,epsf,float,color,subcaption,xcolor}
\usepackage{enumerate}
\usepackage{caption}
\usepackage{natbib}
\usepackage{comment}
\usepackage{subcaption}
\usepackage{url} 
\allowdisplaybreaks
\def\ql{\textcolor{magenta}}

\newcommand{\blind}{0}

\newtheorem{thm}{Theorem}

\theoremstyle{definition} 
\theoremstyle{definition} 
\newcounter{CondCounter}

\newcommand{\bfm}[1]{\ensuremath{\mathbf{#1}}}
\newcommand{\bdm}[1]{\ensuremath{\boldsymbol{#1}}}

\def \a {\bfm{a}}          
\def \d {\bfm{d}}

          \def \u{\bfm{u}}
                           \def \x{\bfm{x}}
\def \y {\bfm{y}}     \def \z {\bfm{z}}

\def \A {\bfm{A}}          
                           
          \def \I{\bfm{I}}
\def \J {\bfm{J}}          
     \def \N {\bfm{N}}     
          \def \R{\bfm{R}}
          
          \def \X{\bfm{X}}
     \def \Z {\bfm{Z}}

\def \alphab   {\bdm{\alpha}}       \def \betab    {\bdm{\beta}}
\def \gammab   {\bdm{\gamma}}       
     
         \def \thetab   {\bdm{\theta}}
        
\def \lambdab  {\bdm{\lambda}}      

\def \Gammab   {\bdm{\Gamma}}

       \def \betas    {{\beta}^{\ast}}

         \def \thetas   {{\theta}^{\ast}}

       \def \betabs    {\bdm{\beta}^{\ast}}
\def \gammabs   {\bdm{\gamma}^{\ast}}       
     
         \def \thetabs   {\bdm{\theta}^{\ast}}

       \def \betah    {\hat{\beta}}

         \def \thetah   {\hat{\theta}}

         \def \thetabh   {\hat{\bdm{\theta}}}

\renewcommand{\hat}{\widehat}

\def \heps     {\hat{\heps}}

\newcommand{\ltwonorm}[1]{\lVert#1\rVert_2}

\newcommand{\supnorm}[1]{\lVert#1\rVert_{\infty}}

\DeclareMathOperator*{\argmin}{argmin}

\def \E        {\mathrm{E}}

\def \var      {\mathrm{Var}}

\def \vec         {\mathrm{vec}}

\def \N           {\mathcal{N}}
\def \R           {\mathcal{R}}

\def \sumj        {\sum_{j=1}^{p_n}}

\def \sumk        {\sum_{k=1}^K}
\def \suml        {\sum_{l=1}^L}
\def \sumt        {\sum_{t=1}^{q_n}}

\def \linear      {\x_{ki}^T\betab+\z_{ki}^T\Gammab \alphab_k}

\def \zero        {\bdm{0}}

\def \thetasone   {\thetab_{S}}

\begin{document}

\def\spacingset#1{\renewcommand{\baselinestretch}%
  {#1}\small\normalsize} \spacingset{1} 


\if0\blind { \title{\bf Modeling Between-Study Heterogeneity for Improved Replicability in Gene Signature Selection and Clinical Prediction} \author{ Naim U. Rashid$^{1,2}$, Quefeng Li$^{1}$, Jen Jen Yeh$^{2,3,4}$, and Joseph G. Ibrahim$^{1}$\\
  }
  \maketitle
\vspace{-3ex}

\begin{center}
  \noindent$^1$Department of Biostatistics, Gillings School of Global Public Health\\

  \noindent$^{2}$Lineberger Comprehensive Cancer Center\\
  
  \noindent$^{3}$ Department of Surgery\\
  
  \noindent$^{4}$ Department of Pharmacology

  \noindent University of North Carolina at Chapel Hill\\
  Chapel Hill, NC, U.S.A.\\\vspace{10pt}

  \noindent Naim U. Rashid naim@unc.edu,\\ 
  \noindent Quefeng  Li quefeng@email.unc.edu,\\
  \noindent Jen Jen Yeh jen\_jen\_yeh@med.unc.edu, and\\  
  \noindent Joseph G. Ibrahim ibrahim@bios.unc.edu \\
\end{center}
 }\fi

\if1\blind
{
  \bigskip
  \bigskip
  \bigskip
  \begin{center}
    {\LARGE\bf Modeling Between-Study Heterogeneity for Improved Replicability in Gene Signature Selection and Clinical Prediction}
  \end{center}
  \medskip
} \fi
\smallskip

\begin{abstract}	
  In the genomic era, the identification of gene signatures associated with disease is of significant interest. Such
  signatures are often used to predict clinical outcomes in new patients and aid clinical decision-making.  However,
  recent studies have shown that gene signatures are often not replicable. This occurrence has practical implications
  regarding the generalizability and clinical applicability of such signatures. To improve replicability, we introduce a
  novel approach to select gene signatures from multiple datasets whose effects are consistently non-zero and account for
  between-study heterogeneity. We build our model upon some rank-based quantities, facilitating integration over different
  genomic datasets. A high dimensional penalized Generalized Linear Mixed Model (pGLMM) is used to select gene signatures
  and address data heterogeneity. We compare our method to some commonly used strategies that select gene signatures
  ignoring between-study heterogeneity. 
  We provide asymptotic results justifying the performance of our method and demonstrate its advantage in the presence of
  heterogeneity through thorough simulation studies.  Lastly, we motivate our method through a case study subtyping
  pancreatic cancer patients from four gene expression studies.
\end{abstract}

\noindent%
{\it Keywords:} Generalized linear mixed models, microarray, penalized likelihood, prediction, RNA-seq.
\vfill

\newpage
\spacingset{1.45} 

\section{Introduction}
In the genomic era, gene signatures are often utilized to subtype cancer patients, determine treatment, and predict
response to therapy \citep{golub1999molecular,swisher2012molecular,sotiriou2007taking}. Such signatures are defined as the
collection of one or more genes whose expression has validated specificity with respect to a particular clinical outcome
\citep{chibon2013cancer}. These signatures are often incorporated into statistical or computational models for predicting
clinical outcome in future patients.  For these reasons, gene signature selection and subsequent clinical prediction is of
significant interest in cancer research.

However, several problems exist with the application of such signatures.  For example, inconsistency in gene signature
selection is common in published biomedical articles.  Gene signatures identified in one article often show little or even
no overlap with the ones identified in another article \citep{waldron2014comparative}. In addition, models based upon
these signatures have shown variable accuracy in predicting outcomes in new clinical studies \citep{sotiriou2007taking,
  waldron2014comparative}, or estimate contradictory effects of individual genes \citep{swisher2012molecular}.  This lack
of replicability presents natural questions towards the generalizability and reliability of utilizing such gene
signatures for clinical prediction \citep{sotiriou2007taking}.

A number of factors contribute to such a lack of replicability. For example, studies with small sample size have been
shown to lack power in selecting gene signatures \citep{sotiriou2007taking} and have low prediction accuracy in new
studies \citep{waldron2014comparative}. Variation in the prevalence of the clinical outcome also affects
replicability. \cite{lusa2007challenges} demonstrate that gene signatures derived from studies with low frequencies of
certain molecular subtypes are less likely to accurately predict molecular subtype in new patients. Study-specific
factors such as variation in laboratory conditions or clinical protocols may also introduce additional variation in the
effects of individual genes.

Differences in data pre-processing is another source. For example, the prediction accuracy of certain classifiers has been shown to be sensitive to the type of normalization method in the pre-processing step \citep{lusa2007challenges, jci2015}. New datasets must be normalized to the training data prior to its application for prediction to correct for technical biases.  However, prior work has shown that this procedure results in ``test-set bias'', where predictions may change due to the samples in the test set or the normalization approach used \citep{patil2015test}. Sophisticated procedures have been developed for microarrays to avoid test-set bias, but still require expression data to come from the same type of microarray chip \citep{mccall2010frozen}. If the new study utilizes a different platform, it is even harder to apply and validate the prediction model. For example, next generation sequencing data measures gene expression on a different scale (positive integer counts) relative to microarray data (continuous measurements). Such a difference typically makes methods developed for one platform not applicable to the other \citep{glas2006converting}.  

To improve replicability, various statistical methods have been developed to integrate data from multiple studies
(horizontal integration) to reach a consensus conclusion.  \cite{richardson2016statistical} give a comprehensive review of
recent developments in this field. Addressing between-study heterogeneity is critical in horizontal data integration, as
data from different studies come from different cohorts, platforms and bio-samples. Several methods
\citep{li2011adaptively,li2014meta} have been developed to account for between-study heterogeneity in horizontal data
integration.  However, these methods mainly focus on variable selection instead of
prediction. 

Motivated by a case study in subtyping pancreatic cancer patients, we develop a new horizontal integration method that
selects gene signatures from multiple datasets and accounts for between-study heterogeneity in variable effects. We apply
a rank-based transformation based upon gene pairs to the raw expression data, facilitating data integration from multiple studies.  We note that some care needs to be taken when merging data from different expression platforms.  More details of this rank-based
transformation will be discussed in Section \ref{sec:methods}. Given the transformed data,
we utilize a penalized Generalized Linear Mixed Model (pGLMM) to select predictors with study-replicable effects and account
for between-study heterogeneity. In particular, we assume the effect of each predictor to be random among different
studies. We design a penalty function to select predictors with nonzero fixed effects in addition to those with non-zero
variance across studies.  We propose to only use predictors with nonzero fixed effects to predict outcome in new subjects,
as their effects are replicable in multiple studies. Through simulation and case studies, we demonstrate that in the
presence of between-study heterogeneity, our proposed method can result in better prediction performance than other
commonly used strategies, especially when the heterogeneity is large. Moreover, as we use the transformed data as
predictors in the pGLMM, our method aims to select gene pairs instead of individual genes for
prediction.

\section{Data}

Pancreatic Ductal Adenocarcinoma (PDAC) remains a lethal disease with a 5-year survival rate of 4$\%$. A key hallmark of
PDAC is the low tumor cellularity of patient samples, which makes capturing precise tumor-specific molecular information
difficult. Due to this fact, genomic subtyping of PDAC to inform treatment selection has been limited.

In a recent study, \cite{moffitt2015virtual} identified genes that are expressed solely in pancreatic tumor cells.  Based
upon these tumor-specific genes, two novel tumor subtypes (`basal-like' and `classical') were identified and validated.
Subtypes were found to be prognostic, in that patients with basal-like tumors had significantly worse median survival than
patients with classical tumors. Lastly, it was found that tumor-specific genes from the basal-like subtype also define a
similar basal-like subtype in breast and bladder cancers, suggesting a common basal-like genomic profile shared across
cancer types. This study represented the largest investigation of primary and metastatic PDAC gene expression thus far and
provided new insights into the molecular composition of PDAC. These insights may be used to make tailored treatment
recommendations.

Given these promising results, methods are needed to robustly predict basal-like subtype. However, existing datasets with
basal-like subtypes in PDAC are limited. Therefore, we utilize the gene expression data from \cite{moffitt2015virtual} in
addition to recently published PDAC RNA-seq data to train a PDAC subtype classifier. Of the three datasets examined in
\cite{moffitt2015virtual}, two are single-channel microarrays (UNC PDAC, UNC Breast Cancer) and one is RNA-seq (TCGA
Bladder Cancer). 
Since the publication of \cite{moffitt2015virtual}, an additional PDAC RNA-seq dataset from The Cancer Genome Atlas (TCGA)
has become available and will also be utilized for training \citep{weinstein2013cancer}.  Expression measurements from
each RNA-seq dataset is summarized in terms of Fragments Per Kilobase of transcript per Million mapped reads (FPKM), a
measurement that accounts for both transcript length and the number of mapped reads within a sample
\citep{trapnell2010transcript}.  This allows for easier comparison of expression measurements across genes and samples
within an RNA-seq study. More modern RNA-seq measurements, such as Transcripts Per Million (TPM, \cite{patro2017salmon})
may also be utilized but were not available from \cite{moffitt2015virtual}. Basic information regarding each dataset is
provided in Table \ref{tab:data}.  Each microarray dataset was normalized as described in \cite{moffitt2015virtual}.

We wish to harness the above datasets to select gene signatures that are predictive of the basal-like subtype. However,
the datasets arise from various expression platforms and therefore have different scales for their expression
measurements.  Furthermore, the datasets have been separately pre-normalized. For these reasons, external validation and
comparison of basal-like subtype prediction models trained separately on each dataset is challenging.  In addition,
integrating datasets to train a single prediction model and select study-consistent variables is difficult, given various
expression platforms and states of pre-processing. The between-study heterogeneity in gene effects may also impact the
selection and estimation of study-consistent variables for subtype prediction.

Motivated by these issues, we propose a novel data integration approach to facilitate between-study comparisons and
merging of samples in Section \ref{sec:methods}.  We also introduce a high dimensional pGLMM to select variables that are
study-consistent while accounting for between-study heterogeneity in their effects.  We compare our method with
several common strategies for gene signature selection and subtype prediction using the data in Table \ref{tab:data}, and
summarize the results in Section \ref{sec:integr-analys-some}.  

\begin{table}[htp]
  \centering
  \footnotesize
  \begin{tabular}{llllll}
    \hline
    Dataset             & Platform   & Sample Size  & Gene Set Size & $\%$ of Basal-like & Pre-normalized? \\\hline
    UNC PDAC            & Microarray & 228          & 19749     & 40$\%$             & Yes            \\
    UNC Breast Cancer   & Microarray & 337          & 17631     & 26$\%$             & Yes            \\
    TCGA Bladder Cancer & RNA-seq    & 223          & 20533     & 47$\%$             & No             \\
    TCGA PDAC           & RNA-seq    & 150          & 20531     & 43$\%$             & No            \\\hline
  \end{tabular}
  \caption{Summaries of four gene expression datasets with basal-like subtype}
  \label{tab:data}
\end{table}

\section{Methods}
\label{sec:methods}
We consider integrating data from $K$ independent studies. For simplicity, we assume there are $n$ subjects in each study
and the total sample size $N = nK$. In the $k$-th study for $k = 1,\ldots,K$, let $\y_k=(y_{k1},\ldots,y_{kn})^T$ be the
vector of $n$ independent responses, $\x_{ki}=(x_{ki,1},\ldots,x_{ki,p_n})^T$ be the $p_n$-dimensional vector of
predictors, and $\X_k=(\x_{k1},\ldots,\x_{kn})^T$. 
Suppose the conditional distribution of $\y_k$ given $\X_k$ belongs to the canonical
exponential family, having the following density function up to an affine transformation that

\begin{equation}
  \label{eq:1} 
  f(\y_k |  \X_k, \alphab_k; \thetab)=\prod_{i=1}^n c(y_{ki})\exp\left[\tau^{-1}\{y_{ki}\vartheta_{ki} - b(\vartheta_{ki}) \} \right],
\end{equation}
where $c(y_{ki})$ is a constant that only depends on $y_{ki}$, $\tau$ is the dispersion parameter, $b(\cdot)$ is a known link function, and the linear predictor
\begin{equation}
  \label{eq:2}
  \vartheta_{ki}=\x_{ki}^T\betab+\z_{ki}^T\Gammab \alphab_k,
\end{equation}
such that $\betab=(\beta_1,\ldots,\beta_{p_n})^T$ is the $p_n$-dimensional vector of fixed effects, $\alphab_k$ is the
$q_n$-dimensional vector of unobservable random effects, $\z_{ki}$ is a $q_n$-dimensional subvector of $\x_{ki}$, and
$\Gammab$ is a lower triangular matrix. We assume $\{\alphab_k \}_{k=1}^K$ are independent and identically distributed
from a general distribution with density $\phi(\alphab_k)$. A common choice of $\phi(\alphab_k)$ is the multivariate
normal distribution $N(\mathbf{0},\mathbf{I}_{q_n \times q_n})$ and
$\Gammab\alphab_k \sim N(\mathbf{0},\Gammab\Gammab^T)$.  In addition, we assume that $\E(\alphab_k)=\zero$ and
$\var(\alphab_k)=\I_{q_n}$. The random component in the linear predictor has $\var(\Gammab
\alphab_k)=\Gammab\Gammab^T$. 
We allow some rows of $\Gammab$ to be identically zero, which implies that the effects of corresponding covariates are
fixed across the $K$ studies. We consider the high dimensional setting for which $p_n\gg n$, $q_n\gg n$, and they both can
grow with $n$. We use the subscript $n$ to denote such a dependence on $n$.

Similar to \cite{chen2003random} and \cite{ibrahim2011fixed}, we reparameterize the linear predictor as
\begin{equation}
  \label{eq:3}
  \vartheta_{ki}=\linear= 
  \begin{pmatrix}
    \x_{ki}^T & (\alphab_k \otimes \z_{ki})^T\J_q
  \end{pmatrix}
  \begin{pmatrix}
    \betab\\
    \gammab
  \end{pmatrix},
\end{equation}
where $\gammab_{t}$ is a $t\times 1$ vector consisting of nonzero elements of the $t$-th row of $\Gammab$, $\gammab=(\gammab_1^T,\ldots,\gammab_{q_n}^T)^T$, and $\J_{q_n}$ is the
$q_n^2\times q_n(q_n+1)/2$ matrix that transforms $\gammab$ to $\vec(\Gammab)$, i.e. $\vec(\Gammab)=\J_{q_n}\gammab$. We define the vector of parameters
$\thetab = (\betab^T, \gammab^T,\tau)^T$ and assume the true value of $\thetab$ is $\thetabs=(\betab^{*T},\gammab^{*T},\tau^{*})^{T}$ such that
$\thetabs=\argmin_{\thetab} \E[-\ell(\thetab)]$, where $\ell(\thetab)$ is the total log-likelihood from the $K$ studies. While the linear predictor $\vartheta_{ki}$ is indeed a function of the parameter $\thetab$, we suppress its dependence on $\thetab$ for the sake of notational simplicity. In addition, we abbreviate $\vartheta_{ki}(\thetabs)$ as $\vartheta_{ki}^{\ast}$, the
value of the linear predictor when the parameters are taken at their true values. As proposed in the above, we would like to identify the set
\begin{equation*}
  S=S_1\cup S_2=\{j: \betas_j\neq 0 \}\cup \{t: \ltwonorm{\gammabs_t} \neq 0 \}.
\end{equation*}
Let $s_{1n}=|\{j:\betas_j\neq0 \}|$ be the cardinality of set $S_1$, $s_{2n}=\sum_{t:\ltwonorm{\gammabs_t}\neq 0} t$ be the cardinality of set $S_2$, $s_n=s_{1n}+s_{2n}$, and
$d_n=p_n+q_n(q_n+1)/2$ be the dimension of the whole problem. In this paper, we consider the case that $d_n$, $p_n$, $q_n$, and $s_n$ change with sample size $n$, but $K$ remains
fixed. 

In order to recover the set $S$, we propose to solve the following penalized likelihood problem:
\begin{equation}
  \label{eq:4}
  \thetabh=\argmin_{\thetab} ~ -\ell(\thetab) + \lambda_1\sumj
  \rho_1(\beta_j)+ \lambda_2\sumt \rho_2(\ltwonorm{\gammab_t}),
\end{equation}
where $\ell(\thetab)=\sumk \ell_k(\thetab)$, $\ell_k(\thetab)$ is the observed log-likelihood from the $k$-th dataset such
that $\ell_k(\thetab) = (1/n)\log\int f(\y_{k}|\X_{k}, \alphab_k; \thetab)\phi(\alphab_k) d\alphab_k$, $\rho_1(t)$ and
$\rho_2(t)$ are some penalty functions, and $\lambda_1$ and $\lambda_2$ are positive tuning parameters. Since (\ref{eq:4})
is a likelihood based method, we may allow the responses $\{\y_k \}_{k=1}^K$ to be of different types. We choose
$\rho_1(t)$ and $\rho_2(t)$ as general folded-concave penalty functions that satisfy condition 8 in Lemma 1 in the
Supplementary Material. Examples of such functions include the $L_1$ penalty, the SCAD penalty \citep{fan2001variable} and
the MCP penalty \citep{zhang2010nearly}. The penalization on $\gammab$ is done in a groupwise manner
\citep{yuan2006model}, namely we regard elements in $\gammab_t$ as a group and penalize its $L_2$-norm. Elements of the
corresponding estimator $\hat{\gammab}_t$ will be either all zero or all nonzero. If $\hat{\gammab}_t=\zero$, the
corresponding variable's effect is regarded as fixed across studies. The selection of such variables (i.e. $S_2$) enables
us to determine which predictors have non-zero fixed effects. We postulate that accounting for study-level heterogeneity
will reduce the bias in fixed effects estimates.

In most applications, we recommend setting $p_n=q_n$ and let the algorithm determine which variables should be regarded as fixed effects. However, if we know that some variables can be treated as fixed effects based on prior knowledge, we only need to impose the penalty $\rho_2$ on the other variables. Based on selections in $S$, we only use predictors with nonzero fixed
effects for prediction. 

Compared to the existing literature on pGLMMs \citep{bondell2010joint,ibrahim2011fixed}, our paper is new in the
following perspectives. First, we deal with a much larger dimension compared to existing articles. In our
application, $p_n$ and $q_n$ can both be greater than 50, yielding at least $2^{100}$ possible models to be chosen from,
whereas the existing articles only consider $p_n = 7$ and $q_n = 3$ in \cite{ibrahim2011fixed} and $p_n = q_n = 16$ in
\cite{bondell2010joint}. In particular, large values of $q_n$ increase the computational complexity of the problem, as the likelihood in
(\ref{eq:4}) involves an integral of dimension $q_n$. To solve such a large-scale problem, a new algorithm is developed to
estimate the pGLMM. More details are given in Section \ref{sec:mcecm-algorithm}. In addition, we give a high-dimensional
asymptotic result in Theorem \ref{thm:1} allowing both $p_n$ and $q_n$ diverge with $n$, while the theory in
\cite{ibrahim2011fixed} requires $p_n$ and $q_n$ to be fixed.

Next, we introduce a technique to facilitate data integration over different studies. The motivation is that even though
the raw values of gene expression may be on different scales in different studies, their relative magnitudes can be
preserved by ranks. Therefore, we propose to use some rank-derived quantities as predictors in models (\ref{eq:1}) and
(\ref{eq:2}), instead of the raw values. We use a variant of the Top Scoring Pair (TSP) approach
\citep{leek2009tspair,patil2015test,afsari2014switchbox}.

Suppose there are $G$ common genes in all $K$ studies. We enumerate $G(G-1)/2$ gene pairs $(g_{ki,s},g_{ki,t})$, where
$g_{ki,s}$ is the raw expression of gene $s$ for subject $i$ in study $k$ and $g_{ki,t}$ is defined similarly. For each
gene pair $(g_{ki,s},g_{ki,t})$, the TSP is an indicator $I(g_{ki,s}>g_{ki,t})$ representing which gene of the two has
higher expression in subject $i$. Such binary indicators are then used as the predictors in (\ref{eq:1}) and (\ref{eq:2}). In other
words, $\x_{ki}$ consists of $G(G-1)/2$ binary variables.  

We view such binary variables as ``biological switches'' indicating how pairs of genes are expressed relative to some
clinical outcome. TSPs were originally proposed in the context of binary classification \citep{afsari2014rank}. We find
that this representation of the original data is also appealing for integrative analysis. First, the TSP
only depends on the ranks of raw gene expression in a sample. Hence, it is invariant to monotone transformations of raw
values. As a result, it is less sensitive to various normalization procedures of data
pre-processing. \citep{afsari2014rank,patil2015test,leek2009tspair}. Second, it simplifies data integration over different
studies. The raw gene expression values may not be directly comparable. After converting them into binary scores, data
from different studies can be pooled together without the need for between-sample or cross-study normalization. Prediction
in new patients is also simplified, as normalizing new patient data to the training set is no longer necessary.

In general, we wish to select gene pairs that are consistent in their relationship with subtypes across multiple studies.
An ideal gene pair is such that one gene in the pair has higher expression than the other gene in one subtype, lower
expression in the other subtype, and has this flip replicated across many subjects.  Each gene in the pair should ideally be differentially expressed between subtypes.
Such ideal gene pairs are less likely to be observed purely due
to technical biases, as this flip in expression is specific to subtype and is also replicated across many subjects.
Indeed, many recent publications utilizing gene pair-based approaches have shown high accuracy and robustness in their
validation datasets, reflecting this point
\citep{afsari2014switchbox,shen2017identification,afsari2014rank,leek2009tspair,kagaris2018auctsp,patil2015test}.

However, some care needs to be taken when merging gene pairs generated from different platforms, especially when merging
microarray data with data from other platforms such as RNA-seq. For microarrays, it is known that differences in absolute
expression between certain genes may not correlate with differences in measured probe-level expression.  Therefore,
merging microarray data with other platforms may reduce the sensitivity to detect such ideal gene pairs.  As a result, our
gene-pair approach is more applicable when data come from the same or similar platforms.  It is also preferable to utilize
more modern expression platforms (such as RNA-seq), as well techniques that correct for GC content and other biases in gene expression measurement \citep{patro2017salmon},
as these approaches may improve the correlation between measured and true expression of genes.  Lastly, our gene pair
approach is predicated on the fact that the genes must also have overlapping expression ranges.  This is commonly observed in our real data application candidate gene set, but may not always be the case.   When the expression ranges of two genes do not overlap,  the corresponding TSP will not flip with respect to subtype across patients, and would therefore would be uninformative for prediction.

\section{MCECM Algorithm}
\label{sec:mcecm-algorithm}
Since the observed likelihood involves intractable integrals, we utilize a Monte Carlo Expectation Conditional
Minimization (MCECM) algorithm for solving (\ref{eq:4}) \citep{garcia2010variable}.  Denote the complete and the observed
data for study $k$ by $\d_{k,c} = (\y_{k},\X_{k}, \alphab_k)$ and $\d_{k,o} = (y_{ki},\x_{ki})$, respectively, and the
entire complete and observed data by $\d_c$ and $\d_o$, respectively.  Let $\lambdab=(\lambda_1,\lambda_2)$. At the $s$-th iteration, given $\thetab^{(s)}$, the E-step is to evaluate the penalized Q-function, given by 

\begin{eqnarray}
  Q_{\lambdab}(\thetab|\thetab^{(s)}) 
  & =& \sum_{k = 1}^{K} E\left\lbrace -\log(f(\d_{k,c}; \thetab|\d_o;\thetab^{(s)}))\right\rbrace +  \lambda_1\sumj \rho_1(\beta_j)+ \lambda_2\sumt \rho_2(\ltwonorm{\gammab_t})\\
  & =& Q_1(\thetab|\thetab^{(s)}) + \lambda_1 \sumj\rho_1(\beta_j)+ \lambda_2\sumt \rho_2(\ltwonorm{\gammab_t}) + Q_2(\thetab^{(s)}),
\end{eqnarray}
where $\d_{k,c} = (\y_k, \X_k, \alphab_k)$, and
$$Q_1(\thetab|\thetab^{(s)}) = -\sumk \int {\log f(\y_{k}|\X_{k}, \alphab_k; \thetab)\phi(\alphab_k | \d_{o,k};\thetab^{(s)})d\alphab_k},$$

$$Q_2(\thetab^{(s)}) = -\sumk \int {\log \phi(\alphab_k)\phi(\alphab_k | \d_{o,k};\thetab^{(s)})d\alphab_k}.$$

Because these integrals are often intractable, we approximate these integrals by taking a Markov Chain Monte Carlo sample
of size $L$ from the density $\phi(\alphab_k | \d_{o,k};\thetab^{(s)})$ using a coordinate-wise metropolis algorithm
described in \cite{mcculloch1997maximum} with standard normal candidate distribution.  This leads to a more efficient
performance for larger $q_n$.  Let $\alphab_k^{(s,l)}$ be the $l$-th simulated value, for $l = 1,\ldots,L$, at the $s$-th iteration of the algorithm. The integral in (6) can be approximated as
$$Q_1(\thetab|\thetab^{(s)}) = -\frac{1}{L} \suml \sumk \log f(\y_k|\X_k, \alphab_k^{(s,l)}; \thetab),$$ $$Q_2(\thetab^{(s)}) = -\frac{1}{L} \suml \sumk \log \phi(\alphab_k^{(s,l)}).$$  The M-step involves minimizing $$Q_{1,\lambdab}(\thetab|\thetab^{(s)}) =
Q_1(\thetab|\thetab^{(s)}) + \lambda_1\sumj\rho_1(\beta_j)+ \lambda_2\sumt \rho_2(\ltwonorm{\gammab_t})$$ with respect to $\thetab=(\betab,\gammab,\tau)$. Minimizing
$Q_{1,\lambdab}(\thetab|\thetab^{(s)})$ with respect to $\tau$ is straightforward and can be done using a standard optimization algorithm, such as the Newton-Raphson Algorithm \citep{rashid2014some}.
Minimizing
$Q_{1,\lambdab}$ with respect to $\betab$ and $\gammab$ is done via the coordinate gradient descent algorithm, leading to more efficient performance in larger dimensions. 

In particular, we utilize three conditional minimization steps.  Prior to minimization, we augment the matrices used in
the linear predictor by ``filling in'' the missing values of $\alphab_k$ with $\alphab_k^{(s,l)}$, repeating the rows of
the original matrices $L$ times and replacing $\alphab_k$ with $\alphab_k^{(s,l)}$ in each of the $L$ repeated rows.  This
leaves us with $\tilde{\Z}_{nKL\times q(q-1)/2} = \left(\tilde{z}_{11}^T,\ldots, \tilde{z}_{nK}^T\right)^T$, where
$\tilde{z}_{ki} = (\tilde{\alphab}_k \otimes \z_{ki})^T\J_q$, and
$\tilde{\alphab}_k = ((\alphab_k^{(s,1)})^T,\ldots,(\alphab_k^{(s,L)})^T)^T$, as well as
$\tilde{\X}_{nKL \times p_n} = (\tilde{x}_{11}^T,\ldots,\tilde{x}_{nK}^T)^T$ to match the dimension of $\tilde{\Z}$, where
$\tilde{x}_{ki} = x_{ki}\mathbf{J}_{L \times 1}$.  We first minimize $Q_{1,\lambdab}$ with respect to $\betab$ given
$\gammab^{(s)}$ and $\tau^{(s)}$ to obtain $\betab^{(s+1)}$ using the coordinate gradient descent approach similar to
\cite{breheny2011coordinate} with predictor matrix $\tilde{\X}$ and offset $\tilde{\Z}\gammab^{(s)}$. We then minimize
$Q_{1,\lambdab}$ with respect to $\gammab$ given $\betab^{(s+1)}$ and $\tau^{(s)}$ to obtain $\gammab^{(s+1)}$ using the
blockwise gradient descent algorithm \citep{breheny2015group} with $\tilde{\X}\betab^{(s+1)}$ serving as an offset.
Therefore, elements of the corresponding estimator $\hat{\gammab}_t$ will be either all zero or all nonzero. If
$\hat{\gammab}_t=\zero$, the $t$-th predictor will be regarded as fixed effect.  By separating the penalized
estimation of $\betab$ and $\gammab$ into two conditional minimization steps, we are able to simplify the variable
selection process into a standard variable selection problem for $\betab$ and a group variable selection problem for
$\gammab$.  Lastly, we minimize $Q_{1,\lambdab}$ with respect to $\tau$ given $\betab^{(s+1)}$ and $\gammab^{(s+1)}$ to
obtain ${\tau}^{(s+1)}$.  This minimization is performed using the Newton-Raphson algorithm.

As $q_n$ increases, the dimension of $\gammab$ also increases.  We utilize an approximation treating the covariance matrix
$\Gammab\Gammab^T$ as a diagonal matrix.  This approach has been demonstrated to be advantageous for high-dimensional
mixed models \citep{fan2012variable}, and also results in greater computational efficiency.  This is because the
accumulative estimation error in estimating the full covariance matrix for large $q_n$ can be much larger than the bias
incurred from utilizing a diagonal covariance matrix.

To ensure that the estimator $\thetabh$ has good properties, the penalty parameter $\lambdab$ has to be appropriately
selected. Two common criteria are generalized cross validation and BIC \citep{wang2007tuning}. However, these criteria
cannot be easily computed in the presence of random effects, because they are functions of the observed likelihood, which
involves intractable integrals. Moreover, it has been shown in \cite{wang2007tuning} that even in the simple linear model,
the generalized cross validation criterion can lead to significant overfitting. Instead, we utilize the ICQ criterion
\citep{ibrahim2011fixed} to select the optimal $\lambdab$ by minimizing

$$ICQ(\lambdab) = −2Q(\hat{\thetab}_{\lambdab}|\hat{\thetab}_0) + c_N(\hat{\thetab}_{\lambdab})$$ where
$c_N(\hat{\thetab}_{\lambdab}) = \text{dim}(\thetab)\times\log(N)$,
$Q(\hat{\thetab}_{\lambdab}|\hat{\thetab}_0) = Q_1(\hat{\thetab}_{\lambdab}|\hat{\thetab}_0) + Q_2(\hat{\thetab}_0)$,
$\hat{\thetab}_0$ is the estimator of $\thetab$ from the full model, and $\hat{\thetab}_{\lambdab}$ is the estimator from
the model fitted with a particular $\lambdab$. As in the EM algorithm, we can draw a set of samples from
$f(\alphab_k | \d_{k,o};\hat{\thetab}_0)$ for $k = 1,\ldots,K$ to estimate $Q(\hat{\thetab}_{\lambdab}|\hat{\thetab}_0)$
for any $\lambdab$.  In higher dimensions, we choose small values for $\lambda_1$ and $\lambda_2$ to approximate
$\hat{\thetab}_0$.  Given the ICQ criterion, we perform a grid search of $(\lambda_1,\lambda_2)$ to find the optimal
values.

For the penalty functions, we consider the MCP penalty for both $\rho_1(t)$ and $\rho_2(t)$, which is defined as
$\rho(t) = \lambda t- {t^2}/{(2\omega)}$ for $t \leq \omega\lambda$ and $\rho(t) = 0$ for $t > \omega\lambda$. Similar to
\cite{breheny2011coordinate}, we choose $\omega = 3$.  Other penalties such as the SCAD and the $L_1$ penalties may be
utilized. Given the promising performance of the MCP penalty in previous publications, we do not explicitly compare between
penalties in this
paper.  

\section{Theory}
\label{sec:theory}
We first introduce some notation. For two sequences $a_n$ and $b_n$, we write $a_n=o(b_n)$ if $a_n/b_n\to 0$;
$a_n \gg b_n$ if $b_n=o(a_n)$; $a_n=O(b_n)$ if $a_n\leq cb_n$ for some positive constant $c$. For a $p$-dimensional vector
$\a$, let $\supnorm{\a}=\max_{1\leq j\leq p} |a_j|$ denote its sup-norm. Let $\a_S$ be a subvector of $\a$ with indices in
the set $S$. For a $p\times p$ matrix $\A$, let $\supnorm{\A}=\max_{1\leq i\leq p} \sum_{j=1}^p |a_{ij}|$ denote the
matrix sup-norm. Denote
$b_n=(\min_{1\leq j \leq p_n}\{|\betas_j| \}\wedge \min_{1\leq t\leq q_n} \{\ltwonorm{\gammabs_t} \})/2$. Let
$\lambda_{ln}=\min\{\lambda_1,\lambda_2 \}$ and $\lambda_{un}=\max\{\lambda_1,\lambda_2 \}$. For simplicity, we assume the
dispersion parameter $\tau=1$ and $\rho_1(t)=\rho_2(t)=\rho(t)$. We define the local concavity of the penalty function as
\begin{equation*}
  \kappa(\rho,\u)=\lim_{\varepsilon\to 0_+}\max_{1\leq j \leq s_n} \sup_{t_1<t_2 \in
    (|u_j|-\varepsilon,|u_j|+\varepsilon)} - \frac{\rho'(t_2)-\rho'(t_1)}{t_2-t_1}.
\end{equation*}
We define a neighborhood of $\thetabs$ as
$\N=\{\thetab=(\betab^T,\gammab^T)^T: \supnorm{\betab_{S_1}-\betabs_{S_1}}\leq c_n, \supnorm{{\gammab}_{S_2}-\gammabs_{S_2}}\leq c_n, \betab_{S_1^c}=\zero, \text{ and }
\gammab_{S_2^c}=\zero\}$, where $c_n=cn^{-\delta}$ for some $c>0$, $0<\delta<1/2$, $S_1^c=\{1,\ldots,p_n \}\backslash S_1$, and $S_2^c=\{1,\ldots,q_n(1+q_n)/2 \}\backslash S_2$.


The main result in Theorem \ref{thm:1} implies that the estimator $\thetabh$ asymptotically recovers $S$ and gives a uniform consistent estimator of $\thetabs_S$. 
\begin{thm}
  \label{thm:1}
  Assume conditions (C1)-(C8) as shown in the Supplementary Material hold.  If $\lambda_{un}\rho'(b_n)=o(n^{-\delta})$,
  $\lambda_{ln}\gg n^{\xi}(s_n^{3/2}b_n/\sqrt{n}+\sqrt{(\log d_n)/n}+s_nn^{-2\delta})$ for $0<\xi<1/2$ and
  $\lambda_{un}\kappa_{0n}=o(\tau_{0n})$, where $\kappa_{0n}=\sup_{\u\in \N_0}\kappa(\rho,\u)$,
  $\N_0=\{\thetab_S\in \R^{s_n}: \supnorm{\thetab_S-\thetabs_S}\leq c_n \}$, and
  $\tau_{0n}=\min_{\thetab\in \N} \lambda_{\min}(\nabla_{\thetasone}^2 \ell(\thetab))$, there exists a sufficiently large
  positive constant $C$ such that with probability greater
  than $1-Ks_nn^{-C}-K(d_n-s_n)d_n^{-C}$, it holds that \\
  (a) $\{j:\thetah_j\neq 0 \}=\{j:\thetas_j\neq 0 \}$.\\
  (b) $\supnorm{\thetabh_S-\thetabs_S}=O(n^{-\delta})$, where $0<\delta<1/2$.
\end{thm}

The convergence rate $\delta$ in statement (b) depends on the minimal signal $b_n$, the dimensionality $d_n$, the sparsity
measurement $s_n$ and the penalty function $\rho(\cdot)$. In general, the larger $b_n$ is and the smaller $d_n$ and $s_n$
are, the faster $\thetabh$ converges. The optimal rate can be as close as a root-$n$
rate.

In Theorem \ref{thm:1}, it is feasible to choose proper tuning parameters $\lambda_1$ and $\lambda_2$ to satisfy all
requirements. For example, if the $L_1$ penalty is used, and we assume $b_n$ is bounded away from 0, we only need to
choose $\lambda_1$ and $\lambda_2$ such that $\lambda_{un}=o(n^{-\delta})$ for some $0<\delta<1/2$ and
$\lambda_{ln}\gg s_n^{3/2}/\sqrt{n}+\sqrt{(\log d_n)/n}$. As long as $s_n=o(\sqrt{n})$ and $\log(d_n)=o(n)$, there exists
a feasible region for $\lambda_1$ and $\lambda_2$. In practice, we tune the optimal $\lambda_1$ and $\lambda_2$ using
methods described in Section \ref{sec:mcecm-algorithm}.

\section{Simulation Studies}
\label{sec:simulation}

\subsection{Oracle setting}
We first examine the oracle setting where the variables relevant to the outcome are known \textit{apriori}. We demonstrate
the performance of our method in comparison to some common strategies to estimate variable effects from multiple
datasets. The first strategy is the traditional study-by-study analysis approach, where variable effects are estimated
separately in each individual study. The second strategy is to combine samples from all studies into a single dataset, and
then estimate variable effects in a single model.  We define a third strategy as a GLMM applied to the merged data,
assuming no penalization on the fixed and random effects. To mimic the process of external validation, we utilize the
fitted model from each strategy to predict outcomes in an externally simulated dataset. The median absolute prediction
error is calculated for each strategy, and is then averaged over
simulations.  
We assess each strategy's performance in terms of the bias of the estimated coefficients as well as the prediction
accuracy under external validation.  We will later examine the variable selection performance under similar conditions
when the set of relevant variables is unknown \textit{apriori}.

Specifically, we generate binary responses representing cancer subtype from a random effects logistic regression model with two predictors and an intercept. A range of sample sizes, number of studies, magnitudes of variable effects, and levels of between-study
heterogeneity are to be inspected. For study $k$, we generate the binary response $y_{ki}$, $i = 1,\ldots,n_k$ such that
$y_{ki} \sim \mathrm{Be}(p_{ki})$ where
$p_{ki}=P(y_{ki}=1|\x_{ki},\z_{ki},\alphab_{k},\betabs )= \exp(\x_{ki}^T\betabs +
\z_{ki}^T\alphab_{k})/\{1+\exp(\x_{ki}^T\betabs + \z_{ki}^T\alphab_{k})\}$, and $\alphab_{k} \sim N_3(0,\sigma^2\I)$,
where $\sigma^2$ controls between-study heterogeneity. To simulate imbalanced sample sizes, we allocate ${N}/{3}$ samples
to study $k = 1$ and evenly distribute the remaining ${2N}/{3}$ samples to the remaining studies.  We perform simulations
for $N = 100, 500$, $K = 2, 5, 10$, $\sigma^2 = 0.5, 1, 2$, $\betabs = (\beta_0^*,\beta_1^*,\beta_2^*)^T = (0, 1, 1)^T$ for moderate predictor effect, and
$\betabs = (0, 2, 2)^T$ for strong predictor effect. For each $k$, we denote the vector of predictors pertaining to subject
$i$ as $\x_{ki} = (1, x_{ki,1}, x_{ki,2})^T$, where we assume $x_{ki,j} \sim N(0,1)$, $j = 1,2$.  We also assume a random intercept and random slope
for each predictor by setting $\z_{ki} = \x_{ki}$.  The external validation set of 100 samples is generated
under the same conditions as the training set to produce $y_{new,i}$ and
$\x_{new,i}$.

For the first strategy (IND), we apply a logistic regression model to each of the $K$ datasets and calculate
$\hat{p}_{new,i}$, the predicted probability of $y_{new,i} = 1$, using $\x_{new,i}$ and the estimated coefficients from
each model. For the second strategy (GLM), we apply a logistic regression model to the merged dataset to obtain
$\hat{p}_{new,i}$. For our method (GLMM), we apply a random effects logistic regression model to the merged dataset to obtain
the estimated fixed effect coefficients, assuming a random slope for each predictor.  Here, only the estimated fixed
effect coefficients are used to obtain $\hat{p}_{new,i}$. In all of the above regression models, we assume the relevant
predictors are known to us and only use them in the model. The median absolute prediction error for each strategy is
calculated as $PE_{med} = \text{median}(|y_{new,i} - \hat{p}_{new,i}|)$, where $i$ varies in the validation set. For the
first strategy, $PE_{med}$ is averaged across the $K$ studies.

\begin{figure}[htp]
  \centering
  \begin{subfigure}{.4\textwidth}
    \centering
    \includegraphics[width=.99\linewidth]{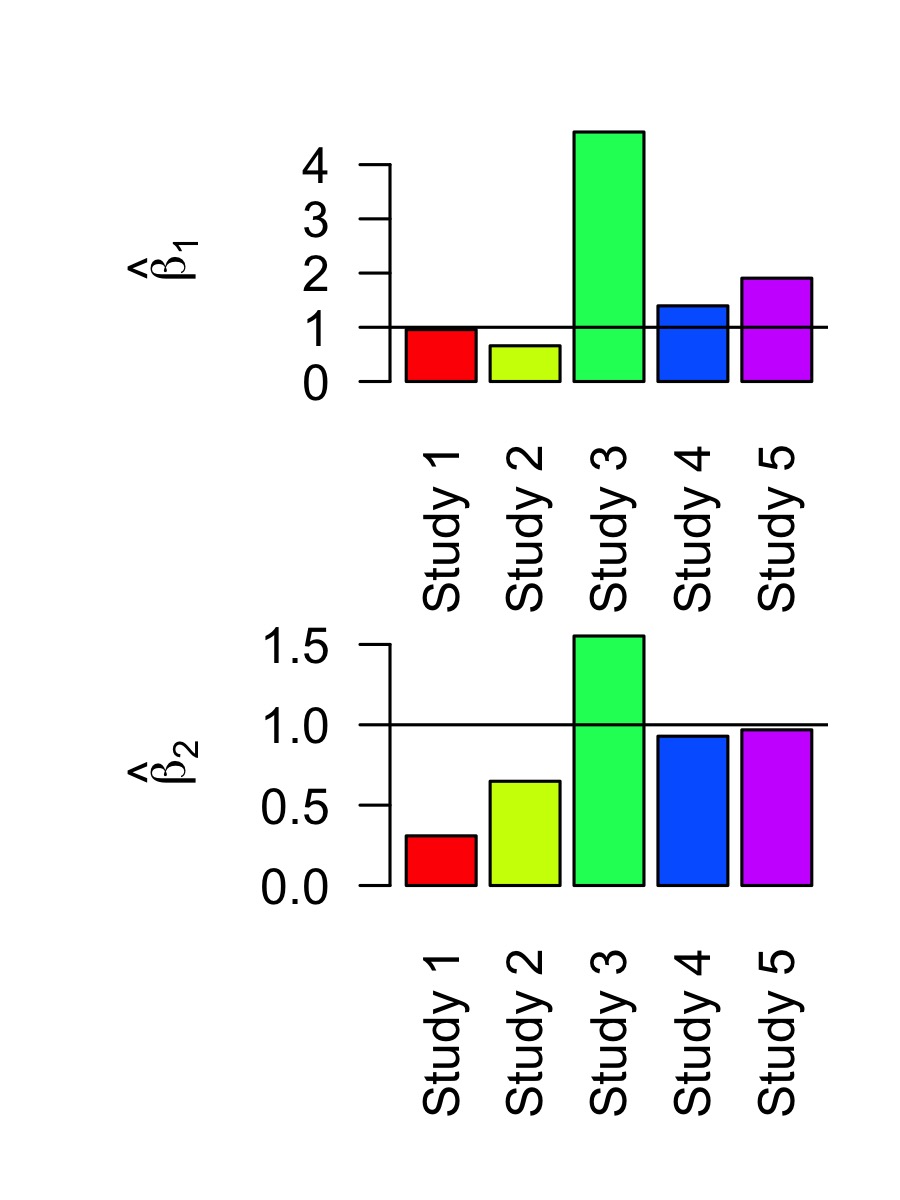}
    \caption{}
    \label{fig:sub1}
  \end{subfigure}%
  \begin{subfigure}{.6\textwidth}
    \centering
    \includegraphics[width=.99\linewidth]{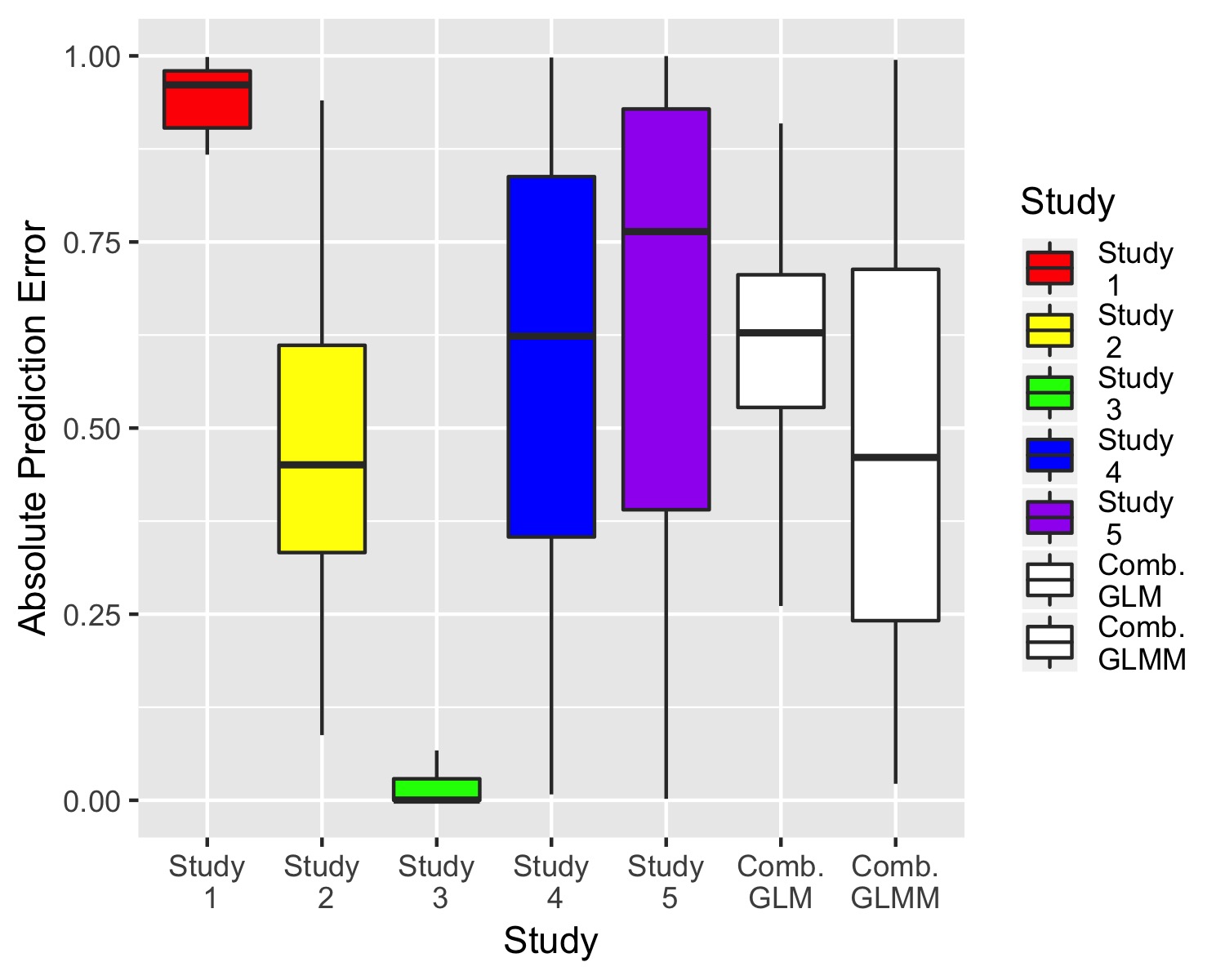}
    \caption{}
    \label{fig:sub2}
  \end{subfigure}
  \caption{Estimation and prediction for strategies 1--3 for a single simulation ($N = 500$, $K = 5$, $\betas_{0}=0$,
    $\betas_{1} = \betas_{2} = 1$, $\sigma^2 = 2$) under the oracle setting. (a) Estimated coefficients in each of the
    five simulated training datasets.  (b) Boxplots of the prediction errors in a simulated external validation
    set. Colored boxplots correspond to the predictions given by the study-by-study
    analysis. 
  }
  \label{fig:test}
\end{figure}

We first illustrate the results of a single simulation in Figure \ref{fig:test}.  In this scenario, we simulate five
studies of a total of 500 samples assuming moderate variable effects and high between-study heterogeneity, i.e., we choose
$N = 500$, $K = 5$, $\betabs = (0, 1, 1)^T$, $\sigma^2 = 2$. Applying the first strategy to the data illustrates the
significant study-to-study variation in the estimated coefficients (Figure \ref{fig:test}, left panel).  This variation is
also observed for the study-level absolute prediction errors in the simulated external validation set (Figure
\ref{fig:test}, right panel).  In this setting, researchers using Study 3 would estimate a strong association between each
predictor and the response, and may further conclude that their model performs well in the validation set.  However,
researchers using Study 1 may conclude otherwise due to the between-study heterogeneity in variable effects.  Combining
data in the second strategy results in smaller prediction errors compared with the first strategy. This observation is in
line with the prior findings suggesting that combining data results in better estimation and prediction
\citep{waldron2014comparative}. However, accounting for heterogeneity further improves the median absolute prediction error.

Our full simulation results are presented in Tables \ref{tab:nopen_beta1} and \ref{tab:nopen_beta2}, where we average
results over 100 simulations per condition.  Several trends are apparent from these results, reflecting our illustration
from Figure \ref{fig:test}.  First, combining data from multiple studies results in an reduction of the median absolute
prediction error ($PE^{GLMM}_{med}$, $PE^{GLM}_{med}$) compared with models trained on individual studies
($PE^{IND}_{med}$); see Table \ref{tab:nopen_beta1}.  We also find that the relative prediction accuracy of the GLMM improves more
when the simulated heterogeneity $\sigma^2$ and the number of studies $K$ increase.  This is due to an increased bias by
the GLM when $\sigma^2$ and $K$ increase. Also, differences in prediction accuracy between the two strategies become more
apparent as the strength of the predictor effects increases (Table \ref{tab:nopen_beta2}). Lastly, the bias of the
estimated coefficients by the GLMM decreases as $K$ and $N$ increase, as more data are available to estimate $\betab$ and
$\Gammab$. In all, combining datasets in strategies two and three leads to better prediction accuracy and accounting for
between-study heterogeneity via our method further improves the performance.

These observations show that even in the oracle setting where the relevant predictors are known, accounting between-study
heterogeneity has important consequences in model estimation and prediction. We assume in our simulations that the
training and validation sets are generated from the same population. We show that even without other complicating factors,
between-study heterogeneity can still impact the accuracy and replicability of common approaches such as strategies one
and two.  While we utilize normally-distributed predictors in our simulations, the impact of between-study heterogeneity
will generally apply to variables from any distribution. In the next section, we show that heterogeneity presents
additional problems in variable selection when important variables are unknown.

\begin{table}[htp]
  \centering
  \tabcolsep=0.2cm
  \renewcommand{\arraystretch}{.75}
  \footnotesize
  \begin{tabular}{rrrrrrrrrr}
    \hline
    $N$ & $K$   & $\sigma^2$ & $\hat{\beta}_1^{GLMM}$ & $\hat{\beta}_2^{GLMM}$ & $\hat{\beta}_{1}^{GLM}$ & $\hat{\beta}_{2}^{GLM}$ & $PE^{GLMM}_{med}$ & $PE^{GLM}_{med}$ & $PE^{IND}_{med}$ \\ 
    \hline
    100  & 2  & 0.5        & 1.03      & 1.06      & 0.90         & 1.03         & 0.33              & 0.34             & 0.39 \\ 
        &   & 1      & 1.11      & 1.06      & 0.84         & 0.81         & 0.38              & 0.40             & 0.43 \\ 
        &   & 2       & 1.01      & 0.97      & 0.76         & 0.49         & 0.42              & 0.43             & 0.46 \\ \hline
        & 5  & 0.5        & 1.14      & 1.15      & 0.95         & 0.93         & 0.34              & 0.35             & 0.39 \\ 
        &   & 1       & 1.12      & 0.98      & 0.77         & 0.74         & 0.40              & 0.42             & 0.43 \\ 
        &   & 2       & 1.22      & 1.06      & 0.53         & 0.49         & 0.45              & 0.47             & 0.48 \\ \hline
        & 10  & 0.5        & 1.15      & 1.20      & 0.93         & 0.96         & 0.33              & 0.35             & 0.39 \\ 
        &   & 1       & 1.07      & 1.01      & 0.73         & 0.67         & 0.38              & 0.41             & 0.43 \\ 
        &   & 2       & 1.02      & 0.87      & 0.40         & 0.40         & 0.43              & 0.47             & 0.47 \\ \hline
    500  & 2  & 0.5        & 1.05      & 1.00      & 1.01         & 0.95         & 0.35              & 0.36             & 0.39 \\ 
        &   & 1       & 0.93      & 1.03      & 0.82         & 0.79         & 0.39              & 0.42             & 0.43 \\ 
        &   & 2       & 0.90      & 0.79      & 0.63         & 0.55         & 0.44              & 0.46             & 0.47 \\ \hline
        & 5  & 0.5        & 0.99      & 1.04      & 0.89         & 0.90         & 0.33              & 0.36             & 0.41 \\ 
        &   & 1       & 0.99      & 0.93      & 0.73         & 0.63         & 0.36              & 0.41             & 0.44 \\ 
        &   & 2       & 0.94      & 0.92      & 0.41         & 0.40         & 0.42              & 0.47             & 0.48 \\ \hline
        & 10 & 0.5        & 0.99      & 1.04      & 0.90         & 0.94         & 0.34              & 0.36             & 0.39 \\ 
        &  & 1       & 1.09      & 0.99      & 0.77         & 0.69         & 0.37              & 0.40             & 0.42 \\ 
        &  & 2       & 0.94      & 0.97      & 0.49         & 0.47         & 0.43              & 0.47             & 0.47 \\ 
    \hline
  \end{tabular}
  \caption{Estimation and prediction under the oracle setting with moderate variable effects for
    $\betabs = (\beta_0^*, \beta_1^*, \beta_2^*)^T = (0, 1, 1)^T$.}
  \label{tab:nopen_beta1}
\end{table}

\begin{table}[htp]
  \centering
  \tabcolsep=0.2cm
  \renewcommand{\arraystretch}{.75}
  \footnotesize
  \begin{tabular}{rrrrrrrrrr}
    \hline
    $N$ & $K$   & $\sigma^2$ & $\hat{\beta}_1^{GLMM}$ & $\hat{\beta}_2^{GLMM}$ & $\hat{\beta}_{1}^{GLM}$ & $\hat{\beta}_{2}^{GLM}$ & $PE^{GLMM}_{med}$ & $PE^{GLM}_{med}$ & $PE^{IND}_{med}$ \\ 
    \hline
    100 & 2 & 0.5 & 2.11 & 2.09 & 1.96 & 1.88 & 0.14 & 0.16 & 0.26 \\ 
        &   & 1 & 2.22 & 2.11 & 1.72 & 1.65 & 0.16 & 0.21 & 0.30 \\ 
        &   & 2 & 1.79 & 2.30 & 1.08 & 1.28 & 0.30 & 0.35 & 0.41 \\ \hline
        & 5 & 0.5 & 2.18 & 2.31 & 1.89 & 1.98 & 0.16 & 0.17 & 0.26 \\ 
        &   & 1 & 2.12 & 2.21 & 1.52 & 1.47 & 0.19 & 0.22 & 0.31 \\ 
        &   & 2 & 1.91 & 1.92 & 0.85 & 0.85 & 0.27 & 0.32 & 0.38 \\ \hline
        & 10& 0.5 & 2.25 & 2.31 & 1.88 & 1.86 & 0.13 & 0.17 & 0.26 \\ 
        &   & 1 & 2.07 & 2.26 & 1.39 & 1.51 & 0.17 & 0.24 & 0.32 \\ 
        &   & 2 & 2.26 & 2.12 & 0.98 & 0.77 & 0.28 & 0.38 & 0.40 \\ \hline
    500 & 2 & 0.5 & 2.04 & 1.98 & 1.97 & 1.93 & 0.15 & 0.17 & 0.26 \\ 
        &   & 1& 1.93 & 1.95 & 1.66 & 1.60 & 0.20 & 0.26 & 0.32 \\ 
        &   & 2 & 2.10 & 1.96 & 1.54 & 1.18 & 0.26 & 0.36 & 0.39 \\ \hline
        & 5 & 0.5 & 2.09 & 2.00 & 1.92 & 1.85 & 0.12 & 0.16 & 0.29 \\ 
        &   & 1& 2.02 & 1.89 & 1.54 & 1.44 & 0.18 & 0.25 & 0.36 \\ 
        &   & 2 & 1.88 & 1.89 & 0.89 & 0.87 & 0.25 & 0.36 & 0.41 \\ \hline
        & 10& 0.5 & 2.01 & 1.98 & 1.85 & 1.85 & 0.15 & 0.17 & 0.26 \\ 
        &   & 1 & 1.93 & 1.91 & 1.41 & 1.40 & 0.18 & 0.25 & 0.31 \\ 
        &   & 2 & 1.81 & 1.83 & 0.88 & 0.90 & 0.27 & 0.36 & 0.40 \\ 
    \hline
  \end{tabular}
  \caption{Estimation and prediction under the oracle setting with strong variable effects for
    $\betabs = (\beta_0^*, \beta_1^*, \beta_2^*)^T = (0, 2, 2)^T$.}
  \label{tab:nopen_beta2}
\end{table}

\subsection{Non-oracle setting }

We again assume that only two variables are relevant to the outcome, but now are unknown \textit{apriori}. We aim to
select these variables from a set of $p$ variables and utilize them to predict outcomes in an external dataset.  In our
simulation, we assume the effects of the remaining $p-2$ variables are zero in all studies.  We simulate our data the same
way as in the previous section, except we now generate $x_{ki,j} \sim N(0,1), j = 1,\ldots p$. We assume
$\x_{ki} = \z_{ki}$. We consider $p = 10$ or 50, $N = 500$, and $K = 5$ or
10.  
Simulation results for these scenarios are given in Tables \ref{tab:pen_beta1} and \ref{tab:pen_beta2}.

We examine three strategies for selecting and estimating the effects of the relevant variables.  For the first strategy
(IND), we apply a penalized logistic regression model separately in each study to select relevant variables. For the second strategy
(GLM), we merge samples from all studies, and then apply the penalized logistic regression to select relevant
variables. Lastly, we apply our method (GLMM) to the merged dataset. The BIC is used to select the optimal tuning
parameters for the first two methods. The optimal tuning parameters of our method are obtained via a grid search based on
the ICQ. In all methods, we choose the MCP penalty. 
Two metrics assessing variable selection performance are presented in Tables \ref{tab:pen_beta1} and \ref{tab:pen_beta2}.
We denote ${TP}$ as the true positives, i.e., the number of correctly selected variables with true non-zero effects; and
${FP}$ as the false positives, i.e., the number of incorrectly selected variables with true zero
effect.

In the low dimensional setting of $p=10$, our method is most advantageous when the heterogeneity is high and the
variables' effects are moderate (Table \ref{tab:pen_beta1}).  In general, strategy two selects fewer true positives but
more false positives compared with our method.  We also find that the first strategy results in the fewest true positives
with the greatest false positives. Its performance worsens when $\sigma^2$ and $K$ increase.  This is due to the smaller
per-study sample size when $K$ increases, as well as the greater chance to have small simulated effects at larger
$\sigma^2$.  Similar to the previous section, we observe that the first two strategies perform worse than our method in
estimation. These results also apply in the high dimensional setting of $p = 50$. In this scenario, the ${FP}^{GLMM}$ is
slightly higher than ${FP}^{GLM}$ in certain settings. But the GLMM has better sensitivity in selecting true positives and
prediction performance. 

Overall, we find that combining datasets improves the variable selection compared with the study-by-study
analysis. 
We also find that accounting for heterogeneity in our method can further improve variable selection, reduce bias, and
reduce prediction error. In the non-oracle setting where the relevant variables are unknown, the prediction errors are
generally larger than the ones in the oracle case. This is due to the uncertainty of variable selection as well as the
bias introduced by
penalization. 

\begin{table}[ht]
  \centering
  \tabcolsep=0.065cm
  \renewcommand{\arraystretch}{.75}
  \scriptsize{
    \begin{tabular}{rrrrrrrrrrrrrrrrr}
      \hline
      $N$ & $p$ & $K$ & $\sigma^2$ & $\betah_1^{GLMM}$ & $\betah_2^{GLMM}$ & $\betah_{1}^{GLM}$ & $\betah_{2}^{GLM}$ & ${TP}^{GLMM}$ & ${FP^{GLMM}}$ & ${TP^{GLM}}$ & ${FP^{GLM}}$ & ${TP^{IND}}$ & ${FP^{IND}}$ & $PE^{GLMM}_{med}$ & $PE^{GLM}_{med}$ & $PE^{IND}_{med}$ \\ 
      \hline
      500 & 10  & 5   & 1          & 0.96              & 1.05              & 0.63               & 0.68               & 1.80          & 0.14          & 1.75         & 0.34         & 0.54         & 1.40         & 0.39              & 0.42             & 0.44 \\ 
          &     &     & 2          & 1.16              & 1.33              & 0.60               & 0.57               & 1.44          & 0.15          & 1.34         & 0.27         & 0.49         & 1.40         & 0.45              & 0.48             & 0.48 \\ 
          &     & 10  & 1          & 0.99              & 0.89              & 0.67               & 0.67               & 1.96          & 0.14          & 1.81         & 0.39         & 0.16         & 1.10         & 0.37              & 0.42             & 0.45 \\ 
          &     &     & 2          & 1.11              & 1.20              & 0.39               & 0.57               & 1.71          & 0.13          & 1.53         & 0.26         & 0.11         & 1.20         & 0.45              & 0.47             & 0.49 \\\hline
      500 & 50  & 5   & 1          & 1.18              & 1.15              & 0.45               & 0.47               & 1.82          & 0.57          & 1.61         & 0.61         & 0.2          & 0.3          & 0.36              & 0.44             & 0.42 \\ 
          &     &     & 2          & 1.12              & 1.18              & 0.55               & 0.44               & 1.47          & 0.91          & 1.12         & 0.42         & 0.23         & 1.4          & 0.36              & 0.43             & 0.44 \\ 
          &     & 10  & 1          & 1.18              & 1.14              & 0.48               & 0.48               & 1.86          & 0.72          & 1.38         & 0.92         & 0.15         & 1.3          & 0.31              & 0.42             & 0.42 \\ 
          &     &     & 2          & 1.23              & 1.38              & 0.55               & 0.53               & 1.51          & 1.08          & 1.23         & 0.4          & 0.13         & 1.3          & 0.36              & 0.41             & 0.43 \\\hline
    \end{tabular}}
  \caption{Variable selection, estimation and prediction under the non-oracle setting with moderate variable effects for
    $\betabs = (\beta_0^*, \beta_1^*, \beta_2^*)^T = (0, 1, 1)^T$.}
  \label{tab:pen_beta1}
\end{table}

\begin{table}[ht]
  \centering
  \tabcolsep=0.065cm
  \renewcommand{\arraystretch}{.75}
  \scriptsize
  \begin{tabular}{rrrrrrrrrrrrrrrrr}
    \hline
    $N$ & $p$ & $K$ & $\sigma^2$ & $\betah_1^{GLMM}$ & $\betah_2^{GLMM}$ & $\betah_{1}^{GLM}$ & $\betah_{2}^{GLM}$ & ${TP^{GLMM}}$ & ${FP^{GLMM}}$ & ${TP^{GLM}}$ & ${FP^{GLM}}$ & ${TP^{IND}}$ & ${FP^{IND}}$ & $PE^{GLMM}_{med}$ & $PE^{GLM}_{med}$ & $PE^{IND}_{med}$ \\ 

    \hline
    500 & 10 & 5 & 1 & 1.94 & 1.93 & 1.48 & 1.45 & 2.00 & 0.07 & 2.00 & 0.11 & 0.40 & 2.00 & 0.19 & 0.25 & 0.33 \\ 
        &   &   & 2 & 2.00 & 2.16 & 1.08 & 1.07 & 1.88 & 0.08 & 1.78 & 0.15 & 0.34 & 2.00 & 0.24 & 0.35 & 0.39 \\ 
        &   & 10 & 1 & 1.90 & 1.90 & 1.42 & 1.36 & 2.00 & 0.08 & 2.00 & 0.10 & 0.34 & 0.80 & 0.18 & 0.25 & 0.39 \\ 
        &   &   & 2 & 1.83 & 2.00 & 0.95 & 0.94 & 1.97 & 0.11 & 1.80 & 0.23 & 0.22 & 0.90 & 0.28 & 0.39 & 0.44 \\\hline 
    500 & 50 &  5 & 1 & 2.19 & 2.04 & 1.48 & 1.53 &    2.00 & 0.84 & 1.58 & 1.62 &    0.00 &   0.00 & 0.18 &  0.3 & 0.37 \\ 
        &   &    & 2 & 2.13 & 1.93 & 1.16 & 0.87 & 1.94 &  2.4 & 1.45 & 1.28 & 0.18 & 1.8 & 0.27 & 0.41 & 0.42 \\ 
        &   & 10 & 1 & 2.09 & 2.16 & 1.46 & 1.49 &    2.00 & 1.36 & 1.28 & 2.84 &  0.3 & 0.2 & 0.16 & 0.34 &  0.4 \\ 
        &   &   & 2 & 2.27 & 2.32 & 0.83 & 0.89 & 1.97 & 1.75 & 1.25 & 2.71 & 0.11 & 1.3 & 0.23 & 0.43 & 0.43 \\ \hline
  \end{tabular}
  \caption{Variable selection, estimation and prediction under the non-oracle setting with strong variable effects for
    $\betabs = (\beta_0^*, \beta_1^*, \beta_2^*)^T = (0, 2, 2)^T$. } 
  \label{tab:pen_beta2}
\end{table}

\section{Improved Clinical Subtype Prediction in Pancreatic Cancer via Horizontal Data Integration}
\label{sec:integr-analys-some}
Using our described data integration approach, we apply four methods to the four datasets described in Table
\ref{tab:data} to predict the `basal-like' subtype in new pancreatic cancer
patients.  
We will show that our method results in better prediction relative to the other methods in the presence of between-study
heterogeneity.


To generate the predictors, we first use $302$ genes that were deemed to be tumor-specific in \cite{moffitt2015virtual}
and appear in all four studies. Then, we apply the rank transformation described in Section 2 in each dataset, enumerating
all possible 45,451 TSPs based on these common genes. To reduce the dimension, we further screen these TSPs by applying a
univariate random effects logistic regression model with respect to each TSP, assuming a random slope and a random intercept. We sort the TSPs  from largest to the smallest by the marginal likelihood from their corresponding random effects logistic regression model.  Then, similar to
\cite{afsari2014switchbox}, we keep TSPs with larger marginal likelihood and remove TSPs sharing one gene with the
higher ranked ones. This reduces potential strong correlation between TSPs sharing same genes (Supplementary Figure 1). After screening, 95 TSPs
remain, of which we select the top 50 ones to be used as covariates in the regression model. We aim to determine the best
subset of the 50 TSPs for prediction.  This results in a total of $2^{50}$ possible fixed effects models and $2^{100}$
possible random effects models.

\begin{figure}[htp]
  \begin{center}
    \includegraphics[width=.8\textwidth]{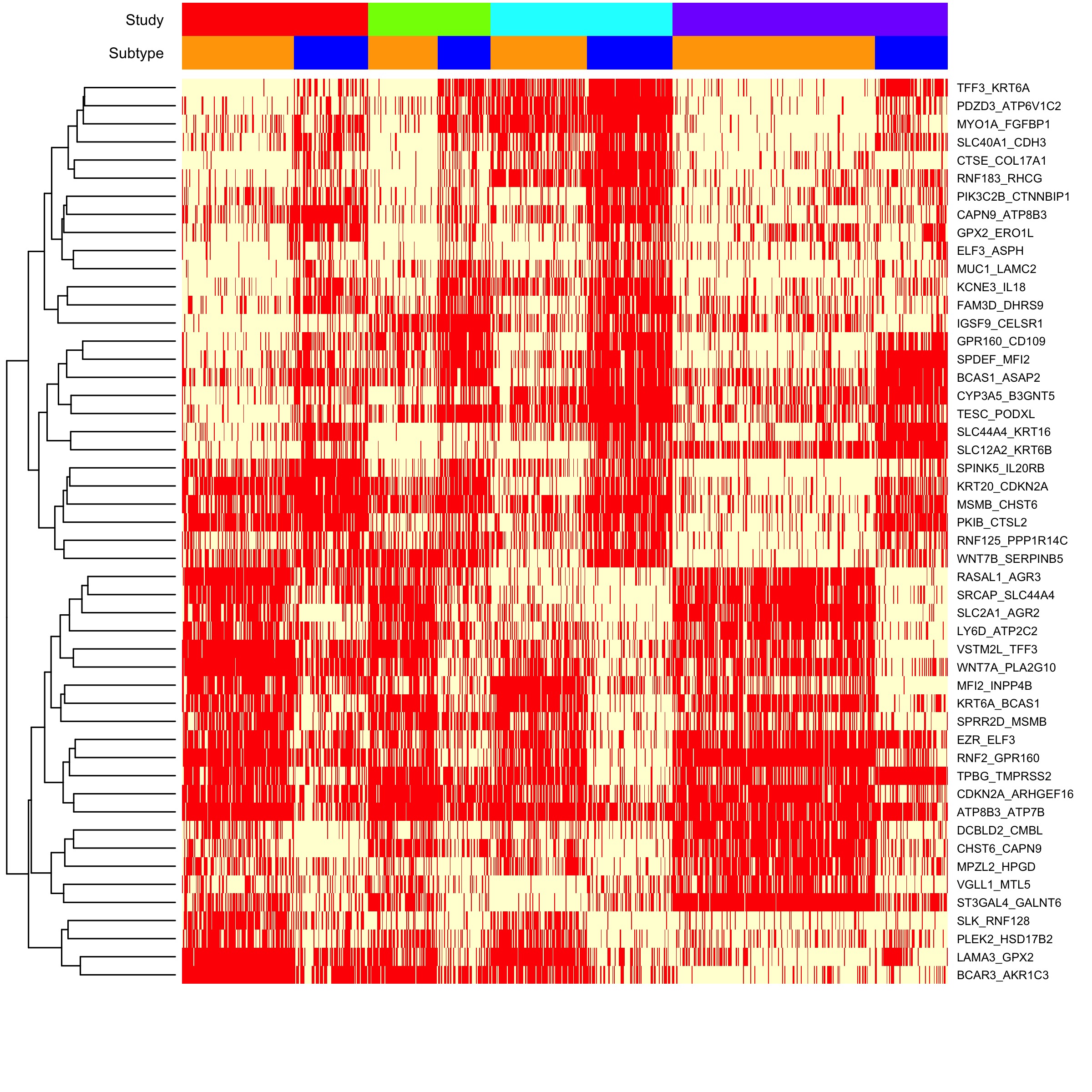}
    \caption{The Matrix of screened TSPs in all studies.  TSPs are labeled in each row as ``A\_B'', where ``A'' indicates
      the name of gene A and ``B'' indicates the name of gene B in the TSP.  Columns indicate samples. Yellow cells in a
      column indicate that the expression of gene A is greater than the expression of gene B, and red cells indicate
      otherwise. The top track (red, green, cyan and purple) indicates study membership. The second track indicates
      patient subtype (blue for basal-like and orange for classical).  Values of TSPs vary significantly across studies,
      where some segregate strongly between basal and classical subtypes in one study but not in other studies.
    }\label{fig:motivating_prob}
  \end{center}
\end{figure}

In Figure \ref{fig:motivating_prob}, we represent the top 50 TSPs for each sample in the four studies.  Yellow cells
indicate that the first gene in the TSP has higher expression than the second gene and the red ones indicate otherwise. It
is clear that certain TSPs have variable association with the subtype across studies, i.e., low replicability. Our goal
is to select the TSPs that are consistently associated with the subtype across studies while accounting for between-study
heterogeneity. 

We compare four methods. For the first method, we apply the penalized logistic regression model (pGLM) to each
dataset. For the second method, we combine all datasets and run the penalized logistic regression model (pGLMC). For the
third method, we run the penalized logistic regression model with random effects on the combined data (pGLMMC). Finally, we
run the Meta-Lasso method \citep{li2014meta} on the combined data. For each subject, we assume the response $y_{ki} = 1$
if the subject is of the basal-like subtype and 0 otherwise. The vector $\x_{ki}$ is the vector of the screened TSPs as
shown in Figure \ref{fig:motivating_prob}. The computational details of the first three methods is the same as described
in the simulation study. For the Meta-Lasso method, the coefficients pertaining to the same TSP in multiple studies are
treated as a group and the composite group penalty is imposed on each group as in \cite{li2014meta}, to select the key
TSPs. The TSPs selected by Meta-Lasso are defined as the ones that have non-zero estimated coefficients in at least one
study. The optimal tuning parameters in Meta-Lasso is determined by the BIC method described in \cite{li2014meta}.

\begin{figure}[htp]
  \begin{center}
    \includegraphics[width=.8\textwidth]{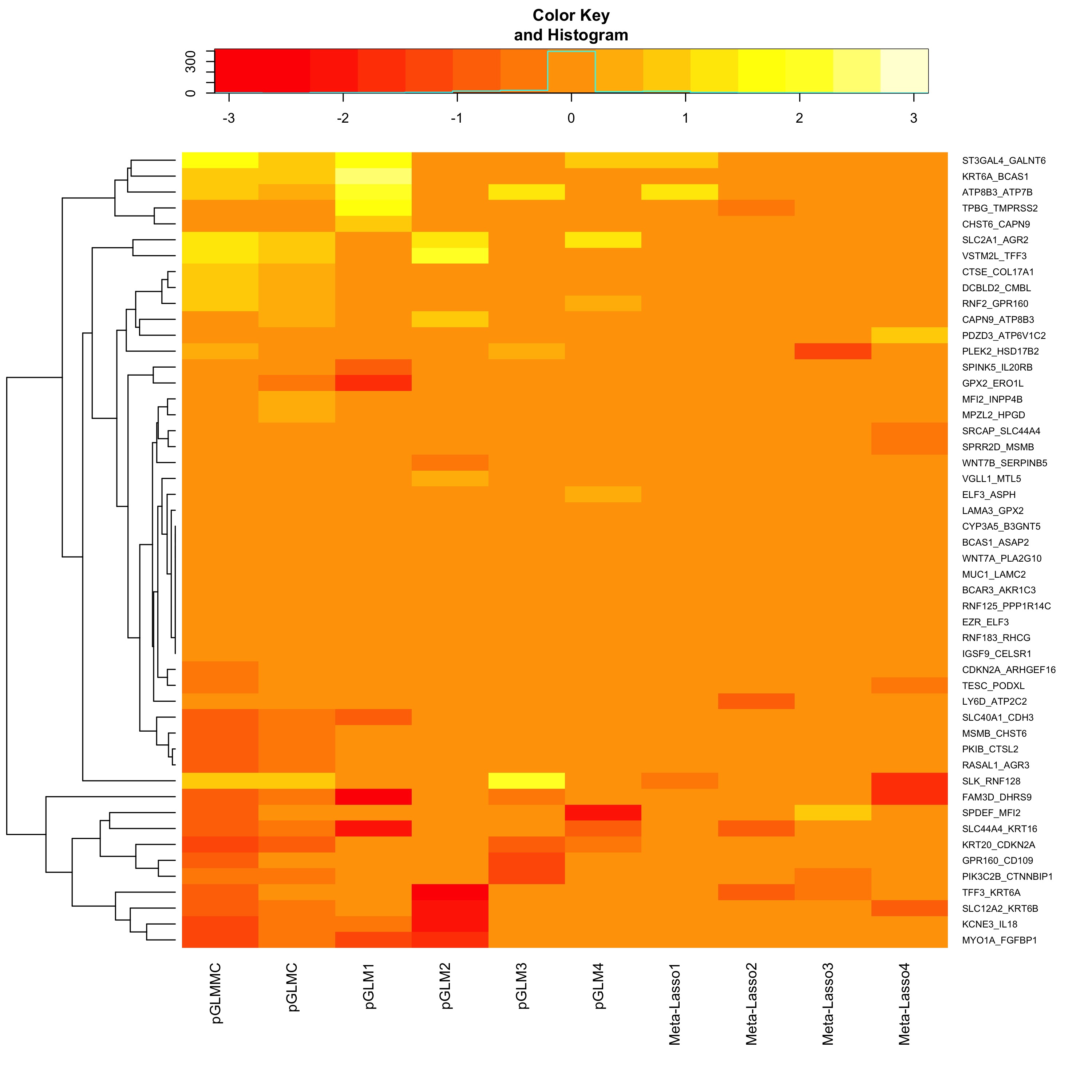}
    \caption{Estimated coefficients given by the four methods. }\label{fig:application_clust}
  \end{center}
\end{figure}

\begin{figure}[htp]
  \begin{center}
    \includegraphics[width=.8\textwidth]{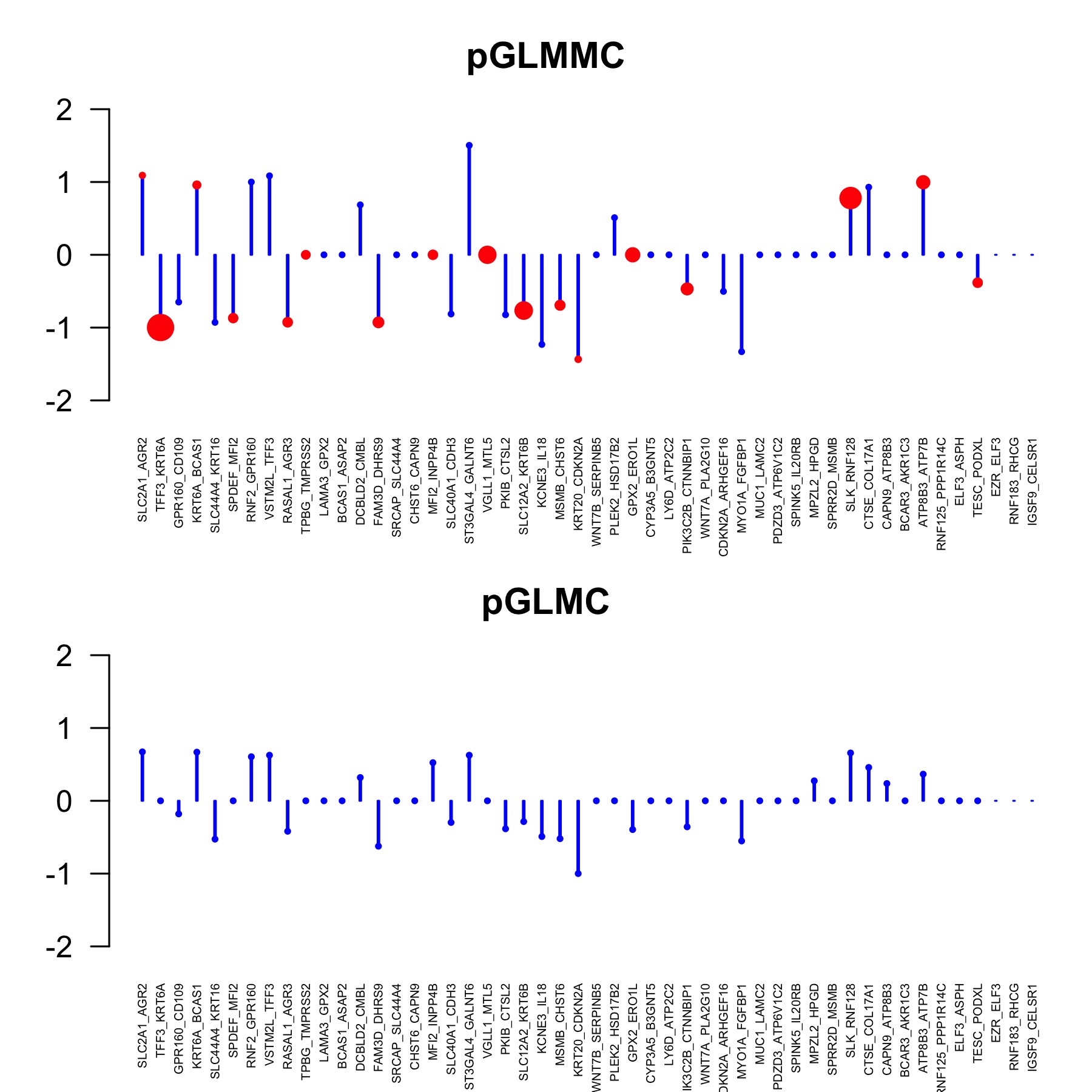}
    \caption{Estimated coefficients given by the pGLMC and the pGLMMC. Red circles indicate variables with non-zero random
      effects estimated by the pGLMMC. Larger red dots indicate larger estimated between-study
      variance.}\label{fig:application_coef}
  \end{center}
\end{figure}

The selected TSPs by the four methods are shown in Figure \ref{fig:application_clust}. Not surprisingly, for the pGLM,
very different TSPs are selected in different studies.  We find that TSPs that are repeatedly selected by the pGLM are
also more likely to be selected by the pGLMC. Our method yields larger estimated coefficients than the pGLMC, especially
for those TSPs selected by both methods (Figure \ref{fig:application_coef}). This mimics findings in our simulation
studies that the estimated coefficients given by the pGLMC are biased in the presence of heterogeneity. Moreover, the
Meta-Lasso selects very different TSPs resulting in poor
replicability.

Next, we evaluate the subtype prediction performance of the four methods. For each method, we hold one dataset out and
train the model using the remaining studies. We utilize this procedure to mimic the process of external validation. For
the pGLM, an ordinary logistic regression model is fitted to each training study using selected TSPs from Figure
\ref{fig:application_clust}. The averages of the three predicted probabilities are assigned to subjects in the holdout
study.  Their absolute prediction errors are then calculated and aggregated from each holdout study. Predictions given by
the Meta-Lasso are done similarly using variables selected by itself. For the pGLMC and the pGLMMC, a single logistic
model is fitted by combining three training datasets and using their own selected TSPs. The predicted probabilities are
then given by such combined models.

\begin{figure}[htp]
  \begin{center}
    \includegraphics[width=.8\textwidth]{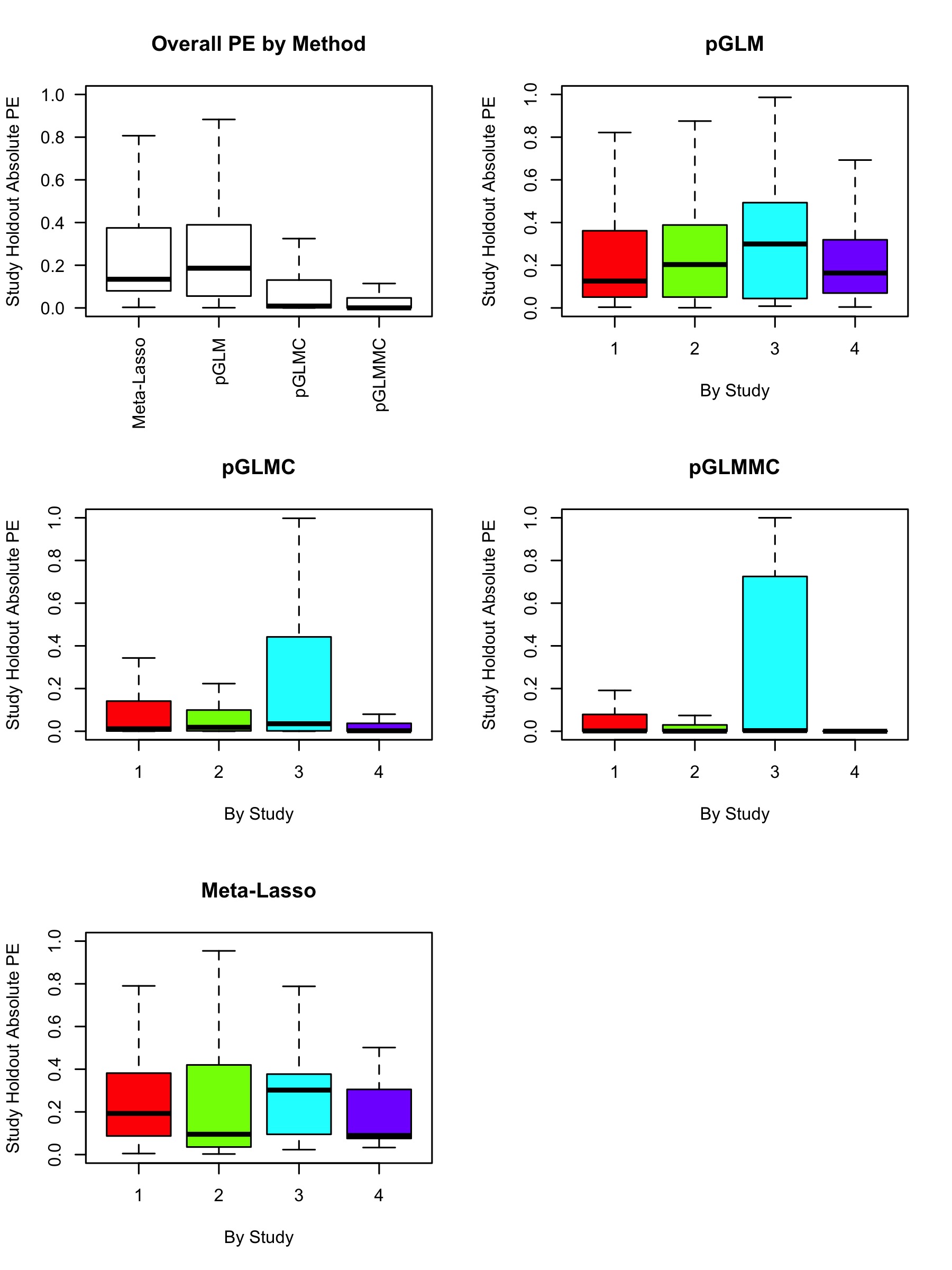}
    \caption{Prediction errors of the holdout studies given by the four methods.}\label{fig:application_holdout}
  \end{center}
\end{figure}

\begin{figure}[htp]
  \begin{center}
    \includegraphics[width=.8\textwidth]{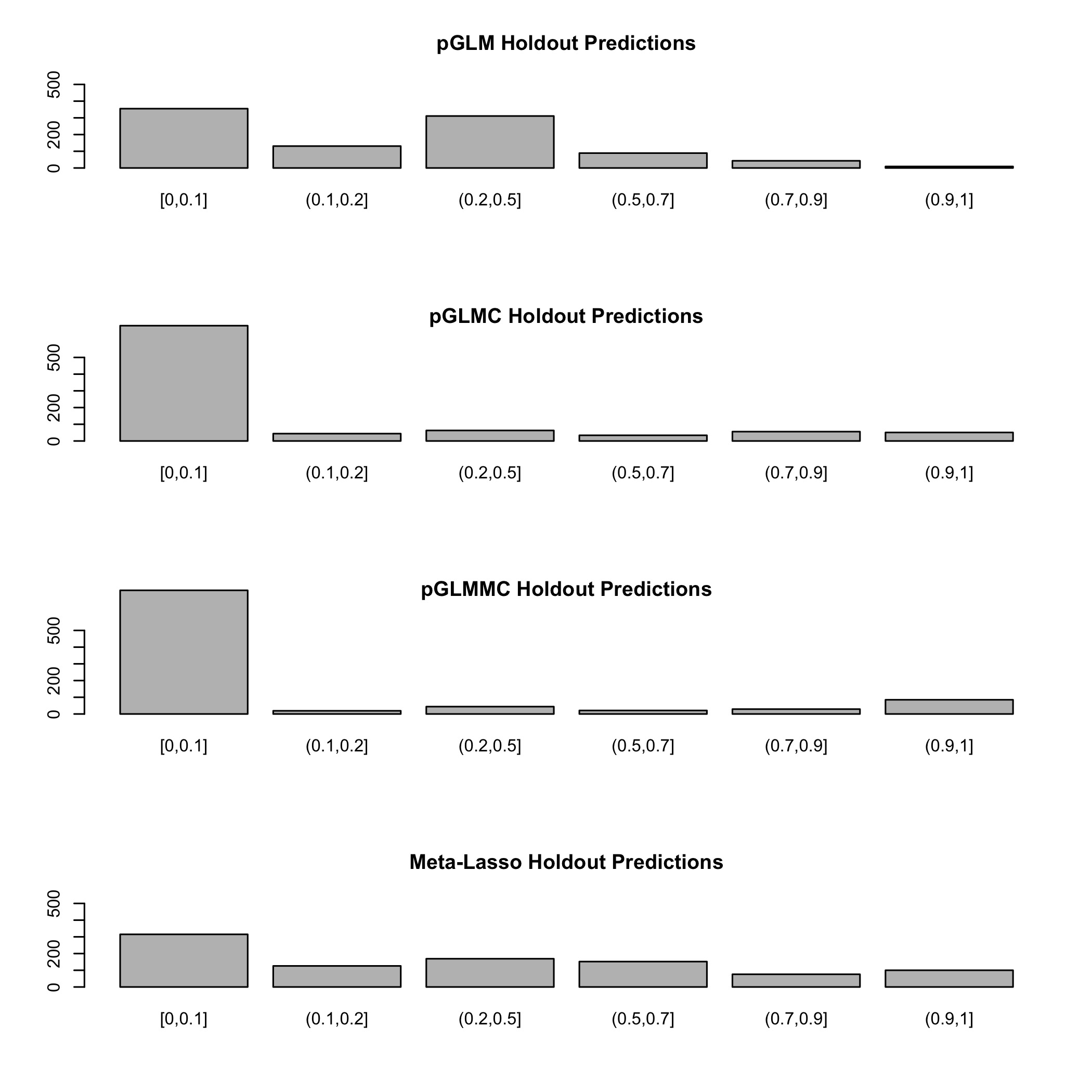}
    \caption{Predicted probabilities of the basal-like subtype given by the four methods.}\label{fig:application_error}
  \end{center}
\end{figure}

Figure \ref{fig:application_holdout} shows the prediction errors given by the four methods in each study. From its top left
panel, we see that the overall performance of the pGLM and the Meta-Lasso is much worse than the pGLMC and the
pGLMMC. These observations reflect the low replicability of predictions from the pGLM and the Meta-Lasso, as the pGLM does not borrow
strength across datasets and the Meta-Lasso is a method mainly focused on variable selection. Similar to our simulation
studies, our proposed pGLMMC method still performs well, despite the variation of its prediction errors on the TCGA
Bladder Cancer dataset is larger than that of the pGLMC.   Its median prediction error however is still the best in this study. In
addition, as shown in Figure \ref{fig:application_error}, our method is more confident than other methods for
classification as most predicted probabilities are either $<10\%$ for $>90\%$. In all, combining datasets significantly
improves the prediction accuracy. By taking heterogeneity into account, our method performs the best out of all
competitors.

In the supplementary material, we provide an alternative screening approach that renders more TSPs and repeat our analysis
therein. Our method's prediction performance is still much better than the pGLM and the Meta-Lasso, albeit it's only
slightly better than the pGLMC (Supplementary Figure 6). This is because the between-study heterogeneity given by the new
screening approach is much smaller than the one shown in this section. Lastly, we also train our method on the microarray
data only and predict on the RNA-seq data, and vice versa. The prediction performance does not change dramatically
(Supplementary Figure 8). 

\section{Discussion}
In this article, we introduce a novel approach accounting for between-study heterogeneity in gene signature selection
and clinical prediction.  We demonstrate through simulations that approaches ignoring existing between-study heterogeneity have
lower prediction accuracy, higher bias, and worse variable selection performance than our method. The common approach of
study-by-study analysis shows the worst performance compared with the integrative approaches. Lastly, we show in a case
study of pancreatic cancer that our method increases prediction accuracy and replicability, where the data integration
is facilitated via a rank-based transformation of the original gene expression
data.  

These results have some important impact. It is often observed that gene signatures derived from individual studies demonstrate
low replicability, even when they pertain to similar clinical outcomes. Our simulation results clearly demonstrate
that this is partially due to the heterogeneity among different studies as small sample sizes in individual studies.  We have
also shown that as the sample sizes of individual studies decreases, the selection sensitivity and prediction performance
also deteriorate.  Selection sensitivity also decreases when the between-study heterogeneity of a gene's effect
increases. On the other hand, combining data from multiple studies improves variable selection and prediction performance
by borrowing strength across studies. However, without taking between-study heterogeneity into account, the naive
combination still performs worse than our proposed method. In the absence of between-study heterogeneity, the random
effects model reduces to the fixed effects model, and therefore we would expect similar performance.  This can be observed
in the additional results in the Supplementary Material. Our simulation and case study results clearly show how the
effects of the same variable may vary significantly between studies, and how this variability impacts prediction. This
explains the lack of replicability observed among published gene signatures.

Finally, we would like to comment that the TSP transformation is one
possible way to enable data integration, and that the choice of the transformation is tangential to the penalized GLMM model that we have proposed.  In addition, the integration of data from
multiple platforms should be taken with care, particularly when merging microarray data with data from other platforms.  Finally, our model aims to select TSPs instead of individual genes.  The success of the TSP transformation relies on the assumption that the raw gene expression has overlapping ranges. Therefore, as pointed out by one reviewer, it could be possible that some genes that are differentially expressed between subtypes will not be selected by our method. 

\bibliographystyle{bibstyle} \bibliography{reference}

\end{document}